**[Title]** Generating automatically labeled data for author name disambiguation: An iterative clustering method

**[Authors]** Jinseok Kim, Jinmo Kim, and Jason Owen-Smith

**[Author Information]**

Jinseok Kim (Corresponding Author)
Institute for Research on Innovation and Science, Survey Research Center, Institute for Social Research, University of Michigan
330 Packard Street, Ann Arbor, MI U.S.A. 48104-2910
734-763-4994|jinseokk@umich.edu|ORCID ID: 0000-0001-6481-2065

Jinmo Kim
School of Information Sciences, University of Illinois at Urbana-Champaign
501 E. Daniel Street, Champaign, IL U.S.A. 61820-6211
jinmok2@illinois.edu

Jason Owen-Smith
Department of Sociology, Institute for Social Research, University of Michigan
330 Packard Street, Ann Arbor, MI U.S.A. 48104-2910
734-936-0463|jdos@umich.edu

Abstract

To train algorithms for supervised author name disambiguation, many studies have relied on hand-labeled truth data that are very laborious to generate. This paper shows that labeled training data can be automatically generated using information features such as email address, coauthor names, and cited references that are available from publication records. For this purpose, high-precision rules for matching name instances on each feature are decided using an external-authority database. Then, selected name instances in target ambiguous data go through the process of pairwise matching based on the rules. Next, they are merged into clusters by a generic entity resolution algorithm. The clustering procedure is repeated over other features until further merging is impossible. Tested on 26,566 instances out of the population of 228K author name instances, this iterative clustering produced accurately labeled data with pairwise F1 = 0.99. The labeled data represented the population data in terms of name ethnicity and co-disambiguating name group size distributions. In addition, trained on the labeled data, machine learning algorithms disambiguated 24K names in test data with performance of pairwise F1 = 0.90 ~ 0.92. Several challenges are discussed for applying this method to resolving author name ambiguity in large-scale scholarly data.

Keywords: author name disambiguation, entity resolution, labeled data, gold standard, supervised machine learning

Introduction

Researchers analyzing scholarly data have faced a common challenge: author names are often ambiguous. For example, many distinct authors may have the same names (homonyms), while an author may use several name variants (synonyms). If name strings are used to identify unique authors, these ambiguous names can lead to misidentification by merging identities associated with homonyms or splitting identities with synonyms.

To date the ambiguity problem has mostly been solved using simple heuristics such as identifying distinct authors by matching their names on forename initials and full surname (Newman, 2001), which has been a dominant practice in bibliometrics for decades (Strotmann & Zhao, 2012). However, name ambiguity can lead this simple approach to produce distorted and sometimes, false positive findings, which has also been well acknowledged by scholars who have relied on the heuristics (Kim & Diesner, 2016).

Computer and information scientists have devised various computational approaches to resolve this problem, showing that supervised machine learning algorithms are promising in disambiguating author names (for a detailed survey, see Ferreira, Gonçalves, & Laender, 2012; Smalheiser & Torvik, 2009). High-performing supervised disambiguation methods tend to be modeled and validated on a few hundreds to thousands of human-labeled cases (for a review on representative hand-labeled data, see Müller, Reitz, & Roy, 2017). There are no general, canonical labeled datasets that can be used across studies (Ferreira, Gonçalves, & Laender, 2012). So, disambiguation scholars usually generate labeled data by hand before training and testing supervised machine learning algorithms.

Generating the hand-labeled data is, however, a daunting task because it requires expensive human coders even for a few thousand name instances. Such labor-intensive methods do not guarantee representativeness or accuracy. For instance, Liu et al. (2014) reported inter-coder disagreement in up to 23% of name instance pairs. As an alternative to manual labeling, some scholars have used the list of name pairs that match on specific criteria such as self-citation relation and shared coauthors, demonstrating that large-scale labeled data can be made automatically (Ferreira, Veloso, Gonçalves, & Laender, 2014; Levin, Krawczyk, Bethard, & Jurafsky, 2012; Torvik & Smalheiser, 2009). Despite their contributions, this matching-based labeling has several known limitations. First, criteria are rarely verified for matching accuracy. Second, performance relies heavily on information availability (e.g., matching on common coauthors may underperform in fields where small teams or sole authorship are the norm). Most importantly, this approach can produce only positive matching pairs of name instances, demanding additional schemes for generating non-matching pairs for training and evaluating algorithmic disambiguation models.

This paper proposes and demonstrates that by synthesizing prior automatic labeling methods, training data for supervised author name disambiguation can be automatically generated by iteratively clustering name instances through the triangulation of metadata and auxiliary information extracted from publication records. Using such automatically labeled data, various supervised machine learning models can be tested for best performance and ambiguity resolution results can be evaluated. In addition, the proposed labeling can be repeated without the added cost of hiring human coders. This can be good news to digital libraries struggling to handle ever-growing, ambiguous bibliographic data. Automatically labeled data can help digital libraries to optimize algorithmic disambiguation models to newly added and updated bibliographic datasets and evaluate their performance on a routine, continuing basis (e.g., every month) at relatively low cost. The following section describes related work to contextualize the proposed method of this paper.

Related Work

Labeled data (also called "gold standard" or "ground truth" data) for author name disambiguation are made up of ambiguous name instances[1] and their associated publication records (such as coauthor names, affiliation, title, venue, publication year, cited references, etc.). A distinct author entity is determined for each name instance using an identification tag (e.g., a unique alphanumeric string). This entity tagging process is called labeling. Depending on how author tags or labels are assigned to name instances, most labeled data can be grouped in three types (Kim, 2018)[2].

The first labeling type is author labels tagged by human coders (e.g., Han, Giles, Zha, Li, & Tsioutsiouliklis, 2004). Typically, this labeling process starts by collating target ambiguous names. Using a digital library or online author profiles, researchers gather ambiguous names based on pre-defined criteria such as names that have the same first forename initial and full surname. Then, the top *k* large groups of names that meet such criteria are selected and publication records related to each name instance are collected. Next, human coders decide which name belongs to whom after comparing each name instance's coauthor name, affiliation, or email address.

This manual process is suited for generating labeled data containing a few hundreds to thousands of name instances. However, hand-labeling is a labor-intensive process even for a small number of names, that is also prone to error due to missing information and inter-coder reliability issues (Han, Zha, & Giles, 2005; Liu et al., 2014; Smalheiser & Torvik, 2009; Song, Kim, & Kim, 2015). Even if human coders reach an agreement on the labeling of certain name instances, their decision can be wrong as shown for the hand-labeled data of Han et al. (2004) (Müller et al., 2017; Santana, Gonçalves, Laender, & Ferreira, 2015; Shin, Kim, Choi, & Kim, 2014)[3]. Moreover, hand-labeled data tend to consist of ambiguous names that are exceptionally difficult to disambiguate (e.g., C. Chen) and, thus may not represent the population of target data in need of disambiguation.

To complement the costly hand-labeled data, some scholars have compared ambiguous author name instances with author profiles registered in other data sources such as authority-controlling digital libraries (e.g., Müller et al., 2017), national researcher profile databases (e.g., D'Angelo, Giuffrida, & Abramo, 2011), and grant data from funding organizations (e.g., Lerchenmueller & Sorenson, 2016).

This data-linkage method can produce labeled data quickly and sometimes at a large scale without human labor. Unlike most hand-labeled data created to train and evaluate disambiguation models, however, the external-authority-based labeling has been utilized mostly for measuring disambiguation performance. Such a limited use is mainly because amounts of linked name instances are decided by coverage of external databases that might be biased toward authors who are grant winners, working in specific nations, or have papers indexed by specific bibliometric services (Lerchenmueller & Sorenson, 2016).

---

[1] This paper distinguishes meanings of author, name, and name instance. An author refers to a distinct entity, a name to a textual string representing the author, and a name instance to an individual occurrence of the name in data. For example, an author (the distinguished professor Mark E. J. Newman at the University of Michigan Department of Physics) can be represented by one or more names (Mark Newman, M. E. J. Newman, etc.) that appear hundreds of times (i.e., instances) through his publication records in bibliometric data.

[2] Other than these three types, a few studies have used synthetic labeled data (e.g., Ferreira, Gonçalves, Almeida, Laender, & Veloso, 2012; Milojević, 2013). Another noticeable labeling approach is to use the intersection set of disambiguation results by multiple algorithms (Vogel, Heise, Draisbach, Lange, & Naumann, 2014)

[3] This does not imply that only Han et al. (2004)'s data contain flaws. No other labeled data than Han et al. (2004)'s have received such intensive scrutiny for errors.

The third type of labeled data have been constructed by generating a list of name pairs that match on a specific identity-matching criterion. Drawing on the observation that authors tend to cite their own papers, for example, some scholars have assumed that if a citing-and-cited pair of papers has the same author name, two instances of the name in each paper indicate the same author identity (for an illustration, see Appendix A). These self-citation name pairs have been used as labeled data usually for evaluating disambiguation results (e.g., Liu et al., 2014; Torvik & Smalheiser, 2009) but sometimes also for training algorithms (e.g., Levin et al., 2012). Other scholars have used email addresses and coauthor names as identity-matching criteria (e.g., Cota, Ferreira, Nascimento, Gonçalves, & Laender, 2010; Ferreira et al., 2014; C. Schulz, Mazloumian, Petersen, Penner, & Helbing, 2014; Torvik & Smalheiser, 2009).

Like the second type of labeled data, this matching-based labeling can automatically produce large-scale, representative labeled data. Unlike the second type, however, this method uses information mostly obtainable in publication records and can, thus, label name instances that are un-linkable using external authority data. Despite such advantages, this approach to automatic labeling still has a room for improvement.

*Problem 1*: Whether matching pairs really represent the same author or not can be uncertain. Although matching accuracy was sometimes validated, for example, via authors' confirmation email (Levin et al., 2012), the common practice of many studies is to presume the accuracy of matching pairs once they meet a pre-defined criterion. An example of incorrect match is the case of two name instances that match on the first-name initial and full surname but have different full first-names (e.g., Mark Newman vs. Mike Newman): they will be decided as a self-citation pair by the common practice using the first-name initial and full surname match for self-citation detection.

*Problem 2*: A second issue is that a criterion can produce different amounts of matching results depending on information availability. For instance, author names from research fields where coauthorship is not prevalent may produce fewer matching pairs than those in areas where team production is a norm.

*Problem 3*: Third and finally, this approach to labeling can produce only true matching pairs for positive training/evaluation sets. In other words, it leaves many true matching pairs undetected and is also unable to identify true non-matching pairs, thus failing to generate negative training/evaluation sets. To address this shortcoming, several studies using this method have devised heuristics (e.g., name pairs different in string *and* sharing no coauthor) to generate non-matching pairs for negative training/evaluation sets, potentially producing trained disambiguation models biased against cases that conform to the negative-matching heuristics but refer to the same authors.

This study synthesizes the second and third types of automatic labeling methods to show that large-scale, representative labeled data can be automatically generated by pairing ambiguous author name instances based on publication metadata and auxiliary information such as self-citation, email addresses, and coauthor names. For this, a set of publication records of computer science articles indexed in the Web of Science (WOS) are selected as a target dataset for author name disambiguation. To improve the accuracy of each identity-matching criterion for names in the WOS data, matching name pairs are compared to author profile information in an external authority source (ORCID) for validation of identity matching (Solution to Problem 1). To increase the amounts of matching pairs, this study triangulates multiple matching criteria to detect matching pairs unfindable by a single criterion (Solution to Problem 2). Most importantly, the triangulation-based method produces clusters of name instances that belong to distinct authors, which can be used to generate true non-matching pairs as well as true matching pairs for training and evaluating disambiguation algorithms (Solution to Problem 3). Details of this automatic labeling process are explained in the following section with the introduction of a real-world example to demonstrate its applicability.

Methodology

*Automatic Labeling Procedure*

*Step1) Finding Feature Matching Rules:* The proposed method for automatic labeling begins by finding the best matching rules for matching features to solve the Problem 1. Specifically, given a dataset of ambiguous names, three information features (email address, coauthor names, and self-citation) which name instance pairs will be matched on are chosen. Then, name instances associated with these features are collected from the dataset. Next, each feature is tested to find a high-accuracy matching rule.

In this study, the matching accuracy of each feature is evaluated using ORCID author profiles. ORCID is an authorship data platform housing publication profiles of more than 5 million authors worldwide. Once registered in ORCID, an author is assigned an ORCID id, which is associated with publication records that are claimed by the author and added by metadata organizations such as Crossref[4] and Europe PMC[5] under the author's authorization (Haak, Fenner, Paglione, Pentz, & Ratner, 2012). For accuracy measurement, each name instance for labeling and its associated publication record is compared to the ORCID author profiles. If a matching author profile is found, its unique ORCID id is assigned to the target name instance. Then, if two name instances judged to be the same by a matching feature are associated with the same ORCID ids, they are regarded as a correct matching case. Linking ORCID ids to name instances in this way allows a high-accuracy matching rule for each feature to be found. Specifically, ratios of correctly matched pairs over the total matched pairs by different matching schemes can be compared to find the best performer.

*Step2) Per-Feature Clustering:* The second step groups name instances into clusters representing distinct authors by applying the high-precision matching rules obtained in the first step. Table 1 illustrates the basic idea of this clustering step with a simplified example.

*Table 1: An Example of Per-Feature Clustering (Before Clustering)*

| Instance No. | Name | Email address |
|---|---|---|
| #1 | Mark Newman | **E1, E2** |
| #2 | M. Newman | E3 |
| #3 | M.E.J. Newman | **E4** |
| #4 | Newman M. | **E5** |
| #5 | M. Newman | E6 |

In the example, five different name instances are related to a matching feature: email address. Initially, each of five instances constitutes a singleton cluster denoted as [#1], [#2], [#3], [#4], and [#5], respectively. Let's assume that Instance #1 and #3 are decided to have the same email address (E1 ≈ E4) according to a matching rule. This email match joins #1 and #3 into a cluster, denoted as [#1|#3], while leaving three singleton clusters ([#2], [#4], and [#5]) intact. Next, let's assume that E2 of Instance #1 is decided by the matching rule to be the same as E5 of Instance #4, which produces another joined cluster [#1|#4]. If two clusters [#1|#3] and [#1|#4] exist, they can be merged into [#1|#3|#4] because Instance #1 appears in both clusters. The result is a newly generated matching pair [#3|#4]. This transitivity closure enables the discovery of additional matches, as presumed in many entity disambiguation studies (Schulz

---

[4] https://www.crossref.org/
[5] https://europepmc.org/

et al., 2014; Whang et al., 2009). As a consequence of such email address matching and transitivity closure, Instance #1, #3, and #4 are assigned the same ids because they belong to the same cluster. Newly assigned cluster ids are shown in Table 2 below (see "Cluster ID" column).

*Table 2: An Example of Per-Feature Clustering (After Clustering)*

| Instance No. | Cluster ID | Name | Email address |
|---|---|---|---|
| #1 | **001** | Mark Newman | E1, E2 |
| #2 | **002** | M. Newman | E3 |
| #3 | **001** | M.E.J. Newman | E4 |
| #4 | **001** | Newman M. | E5 |
| #5 | **002** | M. Newman | E6 |

This per-feature clustering process is described in the pseudo-code below. Here, code lines 1 ~ 10 describe the generation of input data to be processed for feature matching. Specifically, the input *Records* is a list of ids of name instances or clusters with feature information. For example, a name instance consists of an id (#1), a name string (e.g., Mark Newman), an email address (e.g., mejn@umich.edu), coauthor names (e.g., S. H. Strogatz; D. J. Watts), and a list of citing papers (paper1; paper2; paper3, etc.). This information is mapped into a hash table, *recordMap*, for next procedures.

*Algorithm: Pseudo-Code for Per-Feature Clustering*

```
 1:  input: a list Records of instance (or cluster) ids and each id's associated information
 2:  output: a list clusterList of clusters containing author ids that refer to the same author
 3:  recordMap = { }
 4:  for each (id, info) ∈ Records do
 5:      if id ∉ keys(recordMap) then
 6:          recordMap[id] ← info
 7:      else
 8:          recordMap[id] ← <recordMap[id], info>
 9:      end if
10:  end for
11:  clusterList = ( )
12:  for each (i, L_i) ∈ recordMap do
13:      j ← i + 1
14:      for each (j, L_j) ∈ recordMap do
15:          if matchRule(L_i, L_j) = true then
16:              clusterList ← ⟨i, j⟩
17:          end if
18:      end for
19:  end for
20:  repeat
21:      lenList1 ← length of clusterList
22:      for each cluster_i ∈ clusterList do
23:          j ← i + 1
24:          for each cluster_j ∈ clusterList do
25:              if (cluster_i ∩ cluster_j) ≠ ∅ then
26:                  cluster_i ← (cluster_i ∪ cluster_j)
27:                  remove cluster_j from clusterList
28:                  j ← j − 1
```

```
29:             end if
30:          end for
31:       end for
32:       lenList2 ← length of clusterList
33:    until lenList1 = lenList2
34:    return clusterList
```

Lines 11~19 show the matching procedure using the matching function (*matchRule*) decided in Step 1. For example, let's assume that *recordMap* has 5 keys, as in Table 1. The first key ($i$ = #1) is compared to the second key ($j$ = #2) for deciding whether their associated features ($L$; e.g., email addresses) match by *matchRule* (e.g., full string match of pre-@ part for email address). If features are found to match, the pair of $i$ and $j$ is inserted into *clusterList* as [#1|#2]. This process is repeated $j$ = 2, 3, 4, and 5 for $i$ = 1, and $j$= 3, 4, and 5 for $i$ = 2, and so on.

Lines 20 ~34 are implemented for transitivity closure. Given *clusterList* = {[#1|#3], [#1|#4], [#2|#5]}, for example, *cluster₁* [#1|#3] is compared with *cluster₂* [#1|#4] to be merged into *cluster* [#1|#3|#4] because they share #1 (= $cluster_1 \cap cluster_2$). The merged cluster (= $cluster_1 \cup cluster_2$) replaces *cluster₁* [#1|#3] and removes *cluster₂* [#1|#4] from *clusterList*. Now, *cluster₁* [#1|#3|#4] is compared to *cluster₂* [#2|#5][6]. This process is repeated until the length of *clusterList* does not change any more (*lenList1* = *lenList2*). The final output is a list of clusters (*clusterList*), where each cluster represents a distinct author. In the example, two clusters remain: cluster 001 = [#1|#3|#4] and cluster 002 = [#2|#5].

*Step3) Iterative Clustering across Features:* The final step is to repeat the per-feature clustering over other features to address the Problem 2. Table 3 illustrates the situation where each name instance is associated with three information features: email address, self-citation, and coauthor. As a result of Step2 above, five instances are grouped into two clusters (Cluster ID 001 and 002), which are now compared for coauthor match. Let's assume that, per a coauthor-matching rule (e.g., full name string match), Instance #4 in Cluster 001 (= [#1|#3|#4]) and Instance #5 share the same coauthor (C4 ≈ C5). This matching merges #5 into Cluster 001, also attaching [#2] because #2 and #5 belong to the same cluster based on the rule articulated above. Or, Cluster 001(= [#1|#3|#4|#5]) is merged with [#2] as Instance #1 in Cluster 001 is presumed to be in a self-citation relation with Instance #2 (#2 cites #1), thus amalgamating #5, too. This cross-clustering is performed iteratively until no more cluster-merging is possible.

*Table 3: An Example of Iterative Clustering over Multiple Features*

| Instance No. | Cluster ID | Name | Email Address | Coauthor | Self-citation |
|---|---|---|---|---|---|
| #1 | 001 | Mark Newman | E1, E2 | C1 | cites #9 |
| #2 | 002 | M. Newman | E3 | C2 | **cites #1** |
| #3 | 001 | M.E.J. Newman | E4 | C3 | cites #10 |
| #4 | 001 | Newman M. | E5 | **C4** | cites #99 |
| #5 | 002 | M. Newman | E6 | **C5** | cites #11 |

When name instances are merged into clusters across features, feature information associated with a name instance is gathered to be attached to the merged cluster. In Table 3, for example, Instance #1, #3, and #4 were grouped into Cluster 001 through the email-address-based clustering in Step 2, and their associated coauthor information (C1, C3, and C4) is now attached to Cluster 001. When per-feature clustering is conducted over coauthor, matching is performed per *cluster* (with enriched coauthor information), not *instance*, and the aggregate coauthor information is used for matching (code lines 1~10 in Algorithm). This information attachment enables the iterative clustering to detect matching instances that cannot be found by relying on their initially associated information (Ferreira et al., 2014; Whang et al., 2009).

---
[6] Note that the *cluster* [#2|#5] is indexed as $j$ = 2, not $j$ = 3 because the prior merging removes *cluster₂* [#1|#4] from *clusterList*.

After this iterative process produces clusters of name instances, instances in the same cluster are taken to indicate the same author, while instances belonging to different clusters are taken to indicate different authors. This means that true matching pairs of name instances (i.e., positive training sets for machine learning) can be constructed by choosing any two instances from the same cluster and true non-matching pairs (i.e., negative training sets) can be obtained by picking up any two instances from two different clusters. This solves the Problem 3.

*Data and Pre-processing*

*Data*: We apply the proposed automatic labeling to real-world data, "full records" (i.e., including author full names, if available, email addresses, and cited references) of research articles published between 2012 and 2016 in top 100 computer science journals, which were obtained from the Web of Science (WOS)[7]. WOS is frequently used by bibliometric researchers and many disambiguation studies have worked on computer scientist names (Ferreira, Gonçalves, & Laender, 2012). The ranking of journals was based on the Journal Impact Score in 2016 Journal Citation Report[8] for all Computer Science categories. A total of 228,041 name instances were found in 64,991 publication records excluding ones in which author name is null (1 paper) or anonymous (14 papers).

*Email Address*: A total of 154,363 email address instances were found in the downloaded WOS data. As the downloaded data do not tell which email in a paper is associated with what name instance, each email address was matched to an author name automatically. For this, especially, non-alphabetical characters such as dash, dot, and numbers were removed and remaining characters were lower-cased. Then, various combinations of full text string and initials of forename and surname of each name instance (e.g., mejnewman, mnewman, markn, mejn, etc. for Mark E. J. Newman) were compared to the local part (alphabet string before the @ symbol; e.g., 'mejn' in 'mejn@xxxxx.yyy') of email addresses in a paper in which the name instance appears. If two or more name instances were candidates for ownership of an email address, a name instance matched to an email address by one or more full strings was given a priority. If a name instance was matched with two or more emails, the case was excluded from consideration. This matching process associated a total of 140,451 name instances (61.80% of all name instances) with email address instances (one-to-one match). The matching accuracy was 99.2% when evaluated manually on a random sample of 1,000 'email address-name instance' pairs.

*Citation Relations and Coauthorship*: To extract citation relationships among papers, DOIs of papers in "cited references" were compared with those of citing papers. For papers without DOIs, an external dataset[9] recording the paper-level citation relations of 1,568 computer science journals (including most journals in this study's WOS data) was utilized to enhance matching results. A total of 105,051 citation relations among 43,809 papers were found. Generating a coauthor list for an author name instance was straightforward. If three author names (A, B, and C) appear in a paper's byline, each name will have two coauthor names: A's coauthors are B and C, B's are A and C, and C's are A and B.

---

[7] https://clarivate.com/products/web-of-science/web-science-form/web-science-core-collection/
[8] https://clarivate.com/products/journal-citation-reports/
[9] https://static.aminer.org/lab-datasets/citation/dblp.v10.zip

*ORCID-Linkage*: The performance of automatic labeling and name disambiguation is evaluated using ORCID ids linked to name instances as a proxy of ground truth, following Kim (2018). A public data file (released on 10/26/2017) containing 3,564,158 ORCID author profiles in JSON format was obtained.[10] To link author name instances in the downloaded WOS data to ORCID ids, author publication records with DOIs in ORCID data were matched to paper DOIs in the WOS data. Then, a WOS name instance that has the same first-name initial and full surname of the owner author of the matched ORCID record was assigned the author's ORCID id. If two or more name instances are candidates to an ORCID id, they were excluded from linkage. This matching produced a total of 29,386 ORCID id-linked name instances in the WOS data. Among them, 4,945 instances are used to validate the accuracy of matching rules and the iterative clustering performance. The remaining instances (24,441) are set aside as test data.

*Performance Measurement*

A standard evaluation measure for name disambiguation, pairwise F, is used to assess the quality of name instances assignments to authors. A suite of pairwise F metrics − pairwise Precision (*p*P), pairwise Recall (*p*R), and pairwise F1 (*p*F1) − are defined as follows. Note that a name that does not have a comparable pair is not considered for calculation because pairwise F metrics evaluate disambiguation performance at an instance *pair* level.

$$pP = \frac{|\text{name pairs in labeled data} \cap \text{name pairs in ORCID ids data}|}{|\text{name pairs in labeled data}|} \quad (1)$$

$$pR = \frac{|\text{name pairs in labeled data} \cap \text{name pairs in ORCID ids data}|}{|\text{name pairs in ORCID ids data}|} \quad (2)$$

$$pF1 = \frac{2 \times pP \times pR}{pP + pR} \quad (3)$$

Results

*Best Matching Rules for Per-Feature Clustering*

*Email Address Match*: Using the ORCID ids linked to name instances in the WOS data, best matching schemes for email address, self-citation, and coauthorship were found. Three different matching methods were tested for email addresses. First, if two name instances were associated with email addresses sharing the full string format, their ORCID ids were compared to see if they actually refer to the same author. Second, as authors may have multiple email addresses with the same local part (i.e., pre-@) but different domain (i.e., post-@), the accuracy of local part match was also tested. In addition, we also checked whether two email addresses that have the same alphanumeric strings with mechanics (e.g., dots) deleted (for pre-@ part) are associated with the same author. According to the results in Table 4, the full-string-based matching worked best (99.73%) for detecting name instances likely to represent the same author.

*Table 4: Accuracy of Email-Based Identity Matching Methods*

| Matching Scheme | Match Pairs | True Match | Accuracy |
|---|---|---|---|
| Full address | **26,942** | **26,870** | **99.73%** |
| Pre-@ part | 29,706 | 29,081 | 97.90% |
| Alphanumeric character Only | 29,984 | 29,259 | 97.58% |

---

[10] https://figshare.com/articles/ORCID_Public_Data_File_2017/5479792/1

*Self-Citation Match*: Self-citation relationships between two name instances were decided by two different schemes. First, two name instances with the same first forename initial and surname were checked to see if they appear as authors on cited-citing paper pairs. This first-initial-based matching is the common practice of prior studies to decide self-citation name pairs (e.g., Liu et al., 2014; C. Schulz et al., 2014; Torvik & Smalheiser, 2009). As reported in Table 5, the matching accuracy tested on ORCID ids was very high (99.60%). However, the full-name-based matching performed slightly better (99.91%) than the initial-based method. Although initial-based detection produced more matching pairs, the full-string matching approach was chosen to favor high-precision over high-recall because in the clustering stage, incorrectly matched instance pairs can increase the number of incorrectly merged clusters across iterations, which can lead errors to propagate.

*Table 5: Accuracy of Self-Citation Name Instance Pair Detection Methods*

| Matching Scheme | Match Pairs | True Match | Accuracy |
| --- | --- | --- | --- |
| First Initial | 6,035 | 6,011 | 99.60% |
| Full String | **5,513** | **5,508** | **99.91%** |

*Coauthor Match*: Typically, disambiguation studies compare the coauthor names of two ambiguous name instances having the same first forename initial and surname (e.g., Cota et al., 2010; Ferreira et al., 2014; Levin et al., 2012). In addition, coauthor name instances tend to be compared by their first forename initial and surname. Following this convention, this study initialized first forenames of coauthor name instances as well as author name instances before matching. Also, how the number of shared coauthors affects the matching accuracy was tested because several scholars have used different thresholds of coauthor numbers to establish a match (e.g., Ferreira et al., 2014; Levin et al., 2012). The results are presented in Table 6 under the "First Initial" column. As the number of shared coauthors increase, the amounts of pairs to be matched become smaller. But increasing the thresholds improved match accuracy. Besides this initial-based matching, this study tested how using full-strings improves match accuracy. According to the "Full String" column in Table 6, full-string-based matching (for both coauthor and author names) produced smaller amounts of matching pairs with higher accuracy than the initial-based method. Again favoring a high-precision rule to limit error propagation across iterations, we chose full-string matching with a threshold of one coauthor, which produced large numbers of matching pairs (**19,446** > 7,044 > 2,275) with little loss of precision (**99.83%** < 99.86% > 99.82%).

*Table 6: Accuracy of Coauthor-Based Identity Matching Methods*

| Matching Scheme | First-Initial | | | Full String | | |
| --- | --- | --- | --- | --- | --- | --- |
| No. of Shared Coauthors | Match Pairs | True Match | Accuracy | Match Pairs | True Match | Accuracy |
| ≥ 1 | 24,185 | 23,104 | 95.53% | **19,446** | **19,412** | **99.83%** |
| ≥ 2 | 8,112 | 8,038 | 99.09% | 7,044 | 7,034 | 99.86% |
| ≥ 3 | 2,625 | 2,599 | 99.01% | 2,275 | 2,271 | 99.82% |

*Evaluation of Clustering Results*

*Per-Feature Clustering*: Utilizing the matching rules above, name instances associated with email address, self-citation, and coauthor information were clustered using the iterative clustering method explained in the Methodology section. A total of 26,566 name instances (11.69% of all name instances in the downloaded WOS data) that are related to any of the three features were processed for clustering.

Table 7 reports the results when the name instances are clustered based solely on a single feature (by Step 2 Algorithm). The clustering performance was tested on 4,945 ORCID ids linked to the name instances.

*Table 7: Evaluation of Initial Clustering Results Per Feature (Before Iteration)*

| Feature | Number of Clusters | | Pairwise F | | |
|---|---|---|---|---|---|
| | ORCID | Labeled | Precision | Recall | F1 |
| Self-citation | 1,953 | 2,208 | 0.9991 | 0.6945 | 0.8194 |
| Coauthor | 1,953 | 2,585 | 0.9974 | 0.6105 | 0.7574 |
| Email Address | 1,953 | 2,354 | 0.9992 | 0.8279 | 0.9055 |

According to ORCID ids, the name instances should be clustered into 1,953 distinct clusters. In comparison to this truth, the instances clustered only by self-citation resulted in 2,208 clusters, recording a high pairwise precision of 0.9991 but a low pairwise recall of 0.6945. This means that name instances paired by self-citations generally refer to the same authors due to the high-precision matching rule reported in Table 5. However, many name instances that belong to the same authors but are not on self-citing papers failed to be correctly paired as evidenced by low recall. Clustering results by coauthor and email address also show the same pattern of high precision and low recall, implying that clustering based on a single feature is not enough to find all true matching pairs.

*Iterative Clustering*: As the clustering was repeated over other features, the clustering performance increased gradually, as shown in Table 8. For example, the number of clusters decreased from 2,208 (self-citation) to 2,071 (coauthor) and in the end to 1,954 (email address), getting closer to the number of true clusters (1,953). This means that iterative clustering successfully found name instances that belong to the same distinct authors but that were not detected by prior clustering stages. This performance improvement can be confirmed by the recall score which increased incrementally from 0.6945 (self-citation) to 0.8505 (coauthor) and finally to 0.9969 (email address).

*Table 8: Evaluation of Iterative Clustering Results (Incremental in the order of Self-Citation, Coauthor, and Email Address)*

| Feature | Number of clusters | | Pairwise F | | |
|---|---|---|---|---|---|
| | ORCID | Labeled | Precision | Recall | F1 |
| Self-citation | 1,953 | 2,208 | 0.9991 | 0.6945 | 0.8194 |
| + Coauthor | 1,953 | 2,071 | 0.9978 | 0.8505 | 0.9183 |
| + Email Address | 1,953 | 1,954 | 0.9961 | 0.9969 | 0.9965 |

The final results of this iterative clustering procedure were robust to different ordering of features. As illustrated in Table 9, clustering conducted in the order of self-citation, email address, and coauthor matching produced the same final results as the clustering done in the order of self-citation, coauthor, and email address matching. The difference lies in the performance of the middle stage. For example, the number of labeled clusters by email address after self-citation-based clustering was 1,958, which is smaller than 2,071 by the coauthor-based clustering performed after self-citation-based one in Table 8. After an additional clustering iteration on coauthors, the final number of clusters was 1,954, which is the same as the final clustering results in Table 8. The final results were all the same even if the initial clustering started with either coauthor or email address, followed by any clustering order of additional features. A caution is, however, that this is not a natural outcome of the proposed iterative clustering but specific to the case of this study where all name instances are associated with email address, self-citation, and coauthor information. In other words, the iterative clustering may produce different final results on other datasets.

*Table 9: Evaluation of Iterative Clustering Results (Incremental in the order of Self-Citation, Email Address, and Coauthor)*

| Feature | Number of clusters | | Pairwise F | | |
|---|---|---|---|---|---|
| | ORCID | Labeled | Precision | Recall | F1 |
| Self-citation | 1,953 | 2,208 | 0.9991 | 0.6945 | 0.8194 |
| + Email Address | 1,953 | 1,958 | 0.9961 | 0.9934 | 0.9948 |
| + Coauthor | 1,953 | 1,954 | 0.9961 | 0.9969 | 0.9965 |

This re-ordered clustering also shows that one feature can be more useful than others in finding true matching pairs of name instances. For example, the recall gains by email address matching from the baseline result by self-citation-based method were +0.2989 (= 0.9934 – 0.6945), which is larger than +0.1560 (=0.8505 – 0.6945) by coauthor-based clustering applied to the same baseline. This is, however, not unexpected as the email address as a single clustering feature showed the highest recall performance in Table 7.

*Representativeness Checks*

A total of 26,566 instances out of 228,041 author name instances in the downloaded WOS data were labeled as one of 8,218 distinct authors (= clusters) through our iterative clustering process. The size of the resulting labeled data is comparable to that (41,673 instances) of one of the largest hand-labeled datasets for name disambiguation that was manually curated for several months by Korean researchers (KISTI; Kang et al., 2011). Table 10 shows the distribution of name instances per author in the labeled data. As the labeled data in this study consist of name instances that are in self-citation relation with at least one other instance, the minimum number of instances per author is two. Almost 65% of all authors in the labeled data have only two instances. One author has the maximum number of 109 instances that belong to her/him.

*Table 10: Name Instance Distribution per Author*

| No. of Instances | 2 | 3 | 4 | 5 | 6 | 7 | 8 | 9 | 10 ≤ | Total |
|---|---|---|---|---|---|---|---|---|---|---|
| No. of Authors | 5,305 | 1,118 | 685 | 316 | 199 | 143 | 105 | 75 | 37 | 8,218 |
| Ratio (%) | 64.55 | 13.60 | 8.34 | 3.85 | 2.42 | 1.74 | 1.28 | 0.91 | 0.45 | 100.00 |

*Name Ethnicity Distribution*: As the true number of distinct authors in the whole data (of 228,041 name instances) is unknown, it is not clear how these labeled data represent the population data. To address this issue, this study compares the ratios of ethnicity types linked with name instances in the labeled data and the entire dataset. Several disambiguation studies have categorized name instances into groups with different levels of ambiguity based on the findings that some ethnic names are harder to disambiguate than others due to, for instance, common surnames of East Asian authors (e.g., Gomide, Kling, & Figueiredo, 2017; Kim & Diesner, 2016). This grouping has been used to test the sensitivity of disambiguation performance against different types of ethnic names (Lerchenmueller & Sorenson, 2016; Louppe, Al-Natsheh, Susik, & Maguire, 2016).

In this study, an ethnicity tag was assigned to an author name instance by querying its surname to *Ethnea*, an ethnicity classification system (Torvik & Agarwal, 2016)[11]. *Ethnea* assigns a class of ethnicity to a name based on the name's association with its most frequent geo-locations (e.g., "Kim" is most frequently associated with Korea-based institutions), which is weighted by multiclass logistic regression model and probabilistic smoothing (for details see Torvik (2015)). For the case of name instances unseen in the system, "Null" is assigned.

---

[11] http://abel.lis.illinois.edu/cgi-bin/ethnea/search.py. This study uses a batch file of Ethnea for DBLP (2014 version) obtained from http://abel.lis.illinois.edu/cgi-bin/download/request.pl

In Figure 1, ratios of the ten most frequent ethnicities in the whole dataset were compared to those in the labeled data. Although Chinese names are over-represented and English names are under-represented in the labeled data, other name ethnicities are shown to appear in similar proportions with those in the whole data. This means that, at least regarding name derived ethnicities, the labeled data decently represent the whole data.

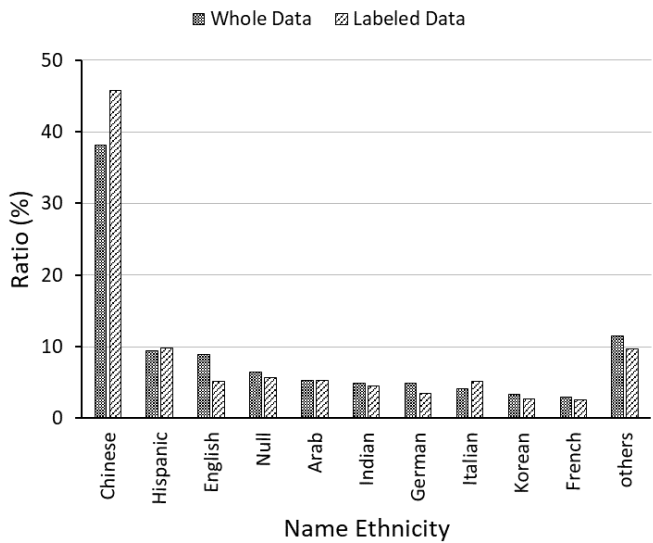

*Figure 1: Ratios of Name Ethnicity in Labeled Data Compared to Whole Data*

*Block Size Distribution*: Another way to see how the labeled dataset represents the whole dataset is to compare their distributions of block sizes. Here, a block size is the number of name instances that match on the first-name initial and full surname. This name grouping has been widely used in disambiguation studies to reduce computation complexity because name instances belonging to different blocks are not compared (e.g., Levin et al., 2012; Louppe et al., 2016; J. Schulz, 2016).

To check representativeness in terms of block size, this study grouped name instances if they share the same initialized forename with a full surname. Next, numbers of blocks with *n* or more instances are counted to calculate their ratios against the total number of blocks. The calculated ratios are then used for comparing block size distributions in the whole and labeled data.

In Figure 2, the ratios of blocks that have *n* or more instances are plotted on a cumulative log-log scale for the cases of the labeled (circles), whole (crosses), and random datasets (triangles). In the whole data, for example, the ratio of blocks with the size of 2 or more is 0.3651 (= 36.51%) of all groups. As depicted in the figure, small-size blocks make up the majority of blocks in the whole data (e.g., blocks with 10 or less make up almost 96%). The labeled data show a similar pattern that small blocks make up the majority. But the plots start on the x-axis value of 2 (and y-axis value of 1) because the smallest blocks in the labeled data contain two instances because each name instance in the labeled data have at least one instance that matches by self-citation relation.

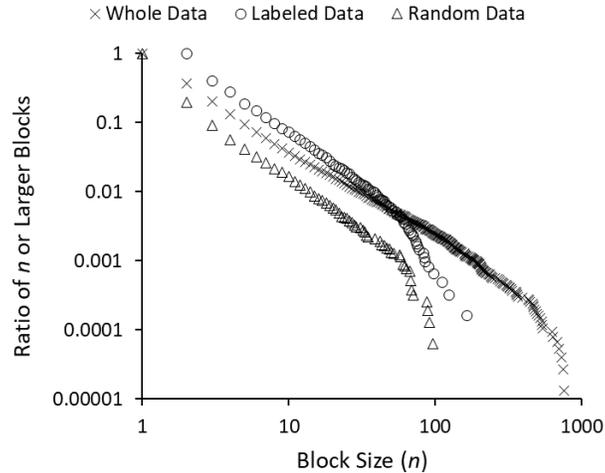

*Figure 2: Cumulative Ratios of Block Size on Log-log Scale*

As observed in Figure 2, the circle plots of the labeled data show a similar trend with the trend for the whole dataset as the value of *n* increases until around 10, when the labeled data trend starts to deviate downward. For a comparison purpose, a subset of the whole data with the same size of the labeled data was randomly generated and its block size distribution (triangles) was depicted on the figure. Slightly different starting points excepted, plots of the labeled and random datasets show a very similar pattern until roughly the size of 60, which constitutes 99.5% of all blocks in the labeled data and 99.92% of all blocks in the random data. These plot trends ($1 \leq n \leq 60$) were fitted to very similar power-law slopes: for the labeled data (-2.523, $R^2 = 0.998$) and the random subset of the entire dataset (-2.557, $R^2 = 0.995$). The two distributions also show a similar downward curvature towards their tails. This implies that the labeled data produced a block size distribution very similar to that of randomly selected instances from the whole data.

*Supervised Disambiguation Using Automatically Labeled Data*

*Training, Development, and Test Data*: To demonstrate the use of automatically labeled data, this study disambiguated 24,441 ORCID-linked name instances (test data) in the whole data by training machine learning algorithms on the automatically labeled data (training data). As a result of the aforesaid clustering iterated over three features, a total of 26,566 author name instances were assigned (=labeled) to 8,218 distinct authors (=clusters). These labeled name instances and their associated information (coauthor and title) are split randomly into two subsets of equal size: the first half as training data to be fed into three commonly used classification algorithms and the second half as development data to optimize thresholds for the hierarchical agglomerative clustering algorithm.

To see how automatically labeled data can contribute distinctively to name disambiguation, the disambiguation results of algorithms trained on them are compared in three ways. First, the same name instances (and associated information) labeled by a single feature – email address, self-citation, and coauthor – are used as baseline training datasets. These baseline datasets are analogous to those used in previous studies that constructed automatically labeled data using each feature: email address (Torvik & Smalheiser, 2009), self-citation (Schulz et al., 2014), and coauthor (Ferreira et al., 2014). Table 11 summarizes the source and characteristics of these automatically labeled data.

*Table 11: Summary of Automatically Labeled Data for Training Algorithms and Test Data (P = Positive, N = Negative)*

| Data Name | Labeling Method | No. of Author Name Instances | No. of Unique Authors | No. of Training Pairs | No. of Development Pairs |
|---|---|---|---|---|---|
| All | Iterative clustering over email, coauthor, and self-citation | 26,566 | 8,218 | P: 19,898 N: 29,070 | P: 20,167 N: 31,944 |
| Email | Per-feature clustering over email | | 10,826 | P: 15,733 N: 34,967 | P: 15,534 N: 34,987 |
| Coauthor | Per-feature clustering over coauthor | | 11,394 | P: 11,894 N: 38,236 | P: 12,688 N: 38,204 |
| SelfCite | Per-feature clustering over self-citation | | 9,436 | P: 13,861 N: 35,615 | P: 15,410 N: 36,062 |
| Test Data | ORCID ids-linkage | 24,441 | 14,936 | P: 28,799 N: 18,107 | |

Second, three hand-labeled datasets in previous studies – AMINER (Wang et al., 2011), KISTI (Kang et al, 2011), and QIAN (Qian et al., 2015) - are used as training and development data to disambiguate name instances in the test data. Table 12 summarizes the source and characteristics of these hand-labeled data. This comparison is based on the idea that if training data created for other disambiguation tasks can produce as much successful disambiguation results as automatically labeled data for the WOS data, generating automatically labeled data for the WOS data would be less meaningful.

*Table 12: Summary of Manually Labeled Data for Training Algorithms (P = Positive, N = Negative)*

| Data Name | Reference | Raw Data Source | No. of Author Name Instances | No. of Unique Authors | No. of Training Pairs | No. of Development Pairs |
|---|---|---|---|---|---|---|
| AMINER | Wang et al. (2011) | DBLP, IEEE, and ACM | 7,528 | 1,546 | P: 61,503 N: 85,427 | P: 63,587 N: 83,231 |
| KISTI | Kang et al. (2011) | DBLP | 41,673 | 6,921 | P: 196,529 N: 303,586 | P: 203,276 N: 290,064 |
| QIAN | Qian et al. (2015) | Multiple datasets | 6,783 | 1,201 | P: 19,871 N: 56,380 | P: 18,156 N: 55,332 |

Third, the iterative clustering proposed for data labeling is applied to the test data using the same matching rules described in the section "Best Matching Rules for Per-Feature Clustering," above. As shown in the section "*Results > Evaluation of Clustering Results*," the iterative clustering produced labeling (= disambiguation) results high in precision, recall, and f1 scores. This implies that the iterative clustering method we propose may be directly applied to disambiguate any test data, possibly eliminating the need of the burdensome machine learning procedure.

*Machine Learning Features*: The feature selection, pre-processing, and similarity calculation described hereafter follows Kim and Kim (2018) and applies to all (automatically and manually) labeled training datasets. Three features – author name, coauthor name and title word – are chosen because they have been used in many author name disambiguation studies. They have been reported to be highly effective in distinguishing author names (Ferreira et al., 2012; Schulz, 2016; Wang et al., 2012). In addition, if many features are used, the effectiveness of labeled data on disambiguation performance cannot be differentiated from that of feature selection.

To pre-process text strings, alphabetical characters were changed into lower-case and encoded into ASCII format. Also, Characters other than alphabets and numbers were replaced by spaces. Commas were, however, left intact because they separate the first-name and surname (last-name) of an author name. After stop-words[12] were deleted, title words were stemmed by the Porter's Stemmer (Porter, 1980)[13]. As a result of this pre-processing, a data instance is formatted as follows: 1(author id)[tab]1(instance id)[tab]kim, jinseok(author name to disambiguate)[tab]kim, jinmo| owen-smith, jason (coauthor names)[tab]automat label data (title words). Similarity between pairs of name instances over each feature was computed by the Term Frequency cosine similarity of 2, 3, and 4-grams (e.g., Han et al., 2005; Kim & Kim, 2018; Levin et al., 2012; Louppe et al., 2016; Santana et al., 2015; Treeratpituk & Giles, 2009). For example, 'jinseok' will converted into a string array of {ji, in, ns, se, eo, ok, jin, ins, nse, seo, eok, jins, inse, nseo, seok}. This is based on the proposition that this n-gram segmentation can be applied consistently across scholar names and title words, contrary to several disambiguation studies that have applied different sets of string comparison rules for names and titles.

*Classification and Clustering*: Pairs of name instances were compared for similarity across three features. Note that comparison was conducted only on names that match on the first-name and full surname (i.e., same block) following the common practice (e.g., Han et al., 2004; Levin et al., 2012; Santana et al., 2015; Wang et al., 2011). Pairwise similarity scores calculated for positive (referring to the same authors) and negative (referring to different authors) pairs constitute training data for three classifiers – Logistic Regression (LR), Naïve Bayes (NB), and Random Forest (RF)[14] – that have been baseline algorithms in many disambiguation studies (e.g., Han, Xu, Zha, & Giles, 2005; Kim & Kim, 2018; Levin et al., 2012; Santana et al., 2015; Torvik & Smalheiser, 2009; Treeratpituk & Giles, 2009; Wang et al., 2012).

Meanwhile, name instances in development and test data were also compared for similarity over each feature by the same procedure applied to training data. Then, disambiguation models by trained algorithms assigned probability scores for the likelihood that two instances represent the same author in each pair in the development and test data. Next, the hierarchical agglomerative clustering algorithm collated name instances of a distinct author based on the pairwise probability scores. Here, a probability score between a pair of name instance represents a similarity distance between the pair. A mean probability score of blocks that maximizes the clustering performance evaluated on the development data[15] was selected as a threshold value in hierarchical clustering algorithms applied to the test data.

*Performance Evaluation*: First, the disambiguation results by three algorithms (LR, NB, and RF) trained on four labeled datasets (Email = labeled by email address matching, Coauthor = labeled by coauthor match, SelfCite = labeled by self-citation match, and All = labeled iteratively over email address, coauthor, and self-citation) were evaluated by pairwise precision, recall, and F1. A set of 24,441 ORCID-linked name instances in the whole data (for details, see Methodology >> Data and Pre-processing >> ORCID Linkage) was used as a proxy of ground truth for evaluation (test data). Figure 3 shows the evaluation results in bar graphs.

---

[12] https://github.com/stanfordnlp/CoreNLP/blob/master/data/edu/stanford/nlp/patterns/surface/stopwords.txt
[13] https://tartarus.org/martin/PorterStemmer/
[14] Classifiers were implemented with parameter settings as follows: L2 Regularization with class weight = 1 (LR), Gaussian Naïve Bayes with maximum likelihood estimator (NB), and 500 trees (after grid search) with Gini Impurity for split quality measure (RF). For more details, see http://scikit-learn.org/stable/index.html
[15] The hierarchical agglomerative clustering algorithm and overall training-test procedure were implemented by modifying codes by Louppe et al. (2016), which are available at https://github.com/glouppe/paper-author-disambiguation

According to Figure 3, algorithms trained on Coauthor and SelfCite scored slightly higher precision than those trained on All (see Figure 3-a). Regarding recall, however, models learned by algorithms from All achieved higher scores than others learned from single-feature-based labeled data (see Figure 3-b). The performance gains in recall by All-based models were so substantial that their harmonic means of precision and recall (i.e., F1 scores; see Figure 3-c) were higher than those of Email, Coauthor, and SelfCite.

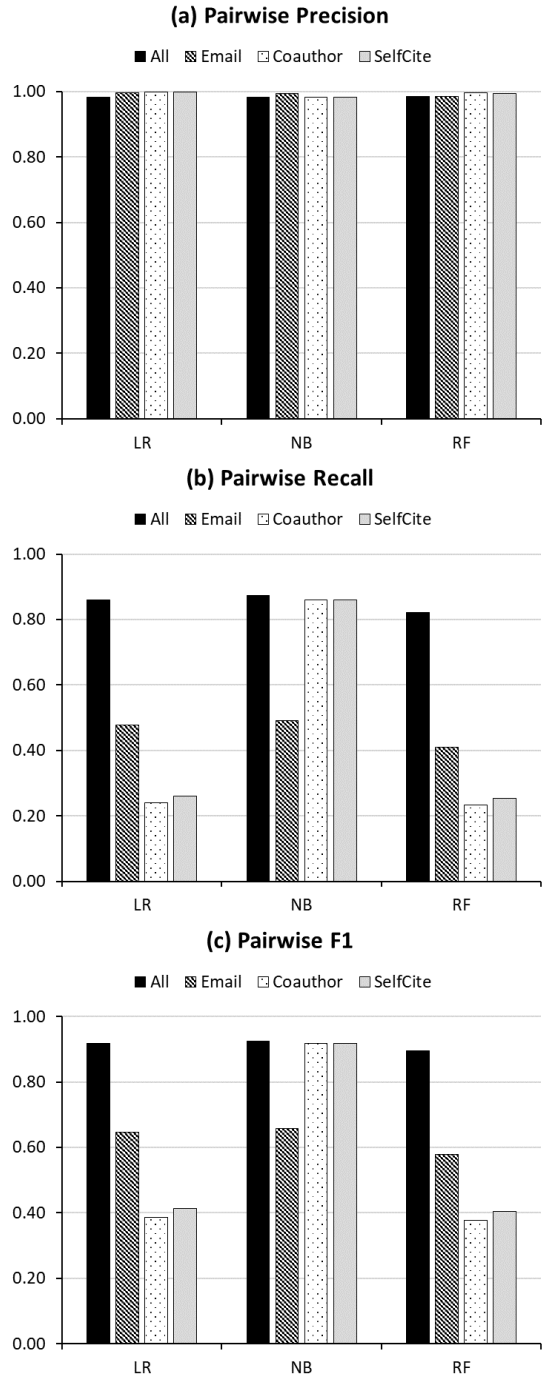

*Figure 3: Evaluation of Disambiguation Results by Three Algorithms Trained on Four Automatically Labeled Data*

These observations align well with the clustering performance reports in Table 7 and 8 where single-feature-based clustering produced higher precisions and lower recalls than those by iterative clustering. This implies that the high-recall labeled data by iterative clustering might affect the high recall on test data by algorithms trained on them. Likewise, the loss of precision by the iterative clustering (possibly due to matching errors) during the generation of labeled data might affect the slightly lower precision of All-trained algorithms on test data than those obtained by the same algorithms trained on baseline labeled datasets with higher precision. Although Naïve Bayes models trained on Coauthor and SelfCite performed quite similarly with the All-trained one, the overall evaluation results indicate that iterative clustering produced labeled data that can improve the performance of disambiguation algorithms.

Next, the disambiguation results by three algorithms trained on four labeled datasets (WOS = automatically labeled from the WOS data; AMINER, KISTI, and QIAN) were evaluated. Figure 4 shows that overall, models trained on automatically labeled data produced better results than those trained on hand-labeled data.

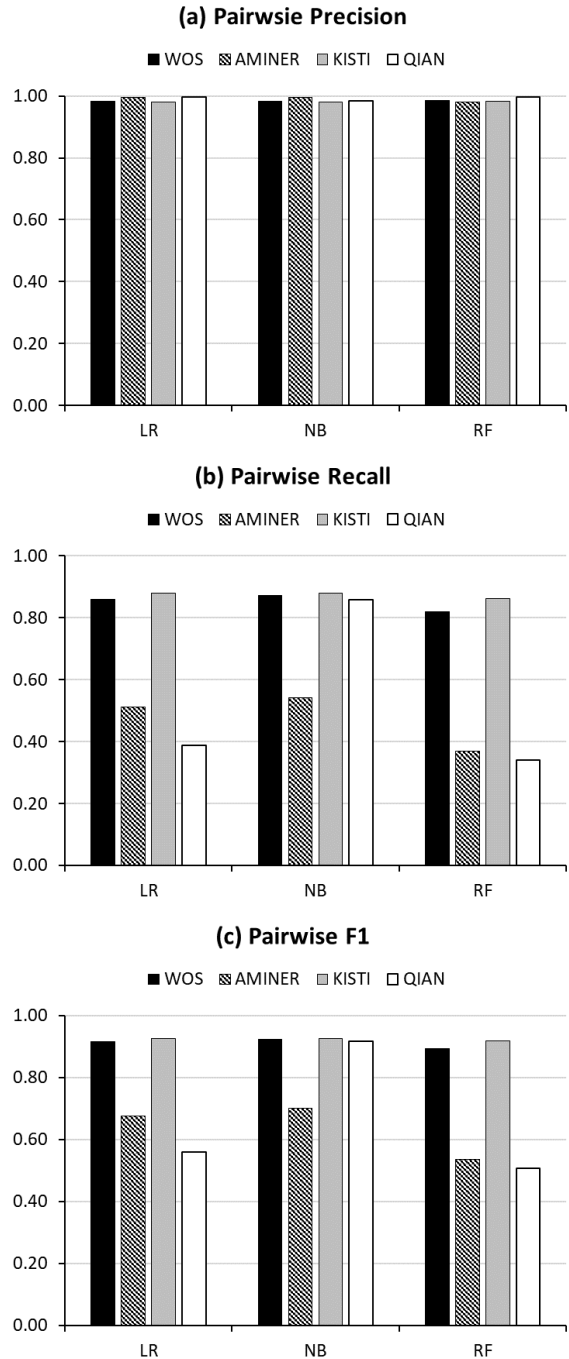

*Figure 4: Evaluation of Disambiguation Results by Three Algorithms Trained on Automatically (WOS) and Manually (AMINER, KISTI, & QIAN) Labeled Datasets*

An interesting observation is, however, that models trained on KISTI (for LR, NB, and RF) and QIAN (for NB) produced quite similar (sometimes slightly better) performances to WOS, while those trained on the other hand-labeled datasets performed worse. Such performances of KISTI and QIAN might be possible in part because the hand-labeled datasets were created from records of publications in computer science. As the name instances of this study were obtained from computer science papers, the hand-labeled datasets might contain the domain-specific characteristics of name distributions (e.g., Chinese names are prevalent; for details, see Figure 1 and Appendix B Figure 7), collaboration pattern, and title term use/frequency that are critical to disambiguating names of computer science scholars. Outside of computer science, however, extensive hand-labeled datasets are rare. Thus, the good performance of manually labeled data may not be replicable in other fields. So, author name disambiguation can get benefits from automatic labeling of training data as proposed in this study.

Finally, the disambiguation results by three algorithms trained on our automatically labeled training data were compared to those by the iterative clustering using three features – email address, self-citation, and coauthor – run on the test data (I-Clustering). Figure 5 reports that the performance by iterative clustering (I-Clustering) is high in precision but low in recall compared to those by the algorithms trained on automatically labeled data. This is not unexpected because iterative clustering works well when feature information is sufficiently complete (e.g., all names have email addresses, self-citation relation, and coauthors as in the automatically labeled data). In the test data, however, the majority of name instances are not associated with one of three features. This indicates that the iterative clustering method has a limitation as a disambiguation method for the test data deficient in feature information.

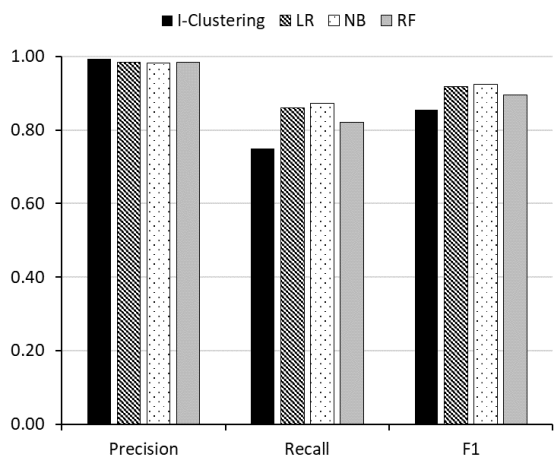

*Figure 5: Evaluation of Disambiguation Results by Three Algorithms (LR = Logistic Regression, NB= Naïve Bayes, and RF = Random Forest) Trained on Automatically Labeled Data (All in Figure 3) in Comparison with Results by Iterative Clustering (I-Clustering)*

Conclusion and Discussion

This study showed that large-scale, representative labeled training data for machine-learning-based author name disambiguation can be generated using publication metadata such as email addresses, coauthor names, and cited references without human curation. Using an external-authority database, high-precision rules for matching name instances could be determined for email address, coauthor names, and self-citation features. Based on these matching rules, name instances were grouped into clusters by a generic entity resolution algorithm able to find matching instance pairs through enhanced feature information and transitivity closure. This clustering was repeated over other features, generating accurately labeled data. The resulting clusters in the labeled data share similar features with the population data in terms of name ethnicity and block size distributions.

In addition, the labeled data were used to train three machine learning algorithms to disambiguate all name instances in the population data with high performance. Models trained on iteratively clustered labeled data improved greatly recall at the slight loss of precision compared to models trained on the same data but labeled by a single-feature-based clustering and external hand-labeled data. This demonstrated that the proposed method can be utilized for studies in need of ad-hoc labeled data to train and test the performance of various disambiguation algorithms. The high performance and scalability of the method has a potential to be applied to supervised machine learning approaches that aim to disambiguate big scholarly data. In addition, such automatically labeled data can be used to evaluate unsupervised machine learning approaches or rule-based methods for author name disambiguation at large scale.

To fully realize this method's potential, however, some issues need to be addressed. First, like other matching-based labeling methods, the performance of the proposed method relies on the availability of matching features. This method may not provide accurately labeled training data for digital libraries that do not record email addresses and cited references. A plausible solution to this problem would be to link other data sources (e.g., AMiner or Microsoft Academic Graph) to the target digital library data to fill missing auxiliary information. Another problem is that as some studies compellingly demonstrate, email addresses are mostly available for recent publications and not all author name instances are associated with them (Levin et al., 2012; Torvik & Smalheiser, 2009). In addition, the number of publications available for disambiguation and the accuracy of their cited references can determine whether self-citation information is rich or relatively scarce. These problems call for an in-depth study about how the imbalance of matching-feature information associated with name instances affects the performance of automatic labeling. Matching feature imbalance is especially critical to expanding the proposed automatic labeling to, for example, a whole dataset in a digital library to obtain representative labeled training data.

Second, as shown in Table 4 ~ 6, even the best matching schemes can produce errors as evidenced by slight decreases in precision with each iteration. Such accuracy decay will impact the performance of iterative clustering because errors propagate in successive stages. This implies that a better understanding of error propagation in iterative clustering algorithms is necessary before applying this study's method to labeling data involving many iterations. Together, these two problems indicate that automatic labeling can be improved by expanding accurate and publicly available matching features beyond the three tested in this study. Affiliation information in publication records or multiple sources of researcher ids in other digital libraries may enhance automatic labeling efforts.

Finally, different coverage in target data can lead to challenges for validation of matching rules and measurement of labeling accuracy. As an evaluation source, this study relied on ORCID ids which is author managed and covers a wide range of scientific domains. The high accuracy of ORCID's author profiling was confirmed for more than 700,000 name instances in DBLP associated with ORCID ids[16], and, for that reason, ORCID data were used for evaluating name disambiguation performance of DBLP (Kim, 2018). But its accuracy for other domains than computer science has not properly evaluated. Another important limitation is that ORCID records may not cover all publications of an author because individual authors decide the entry and update of their publication information in the ORCID system. Also, ORCID ids-linked authors in our data over-represent Hispanic authors while they under-represent Chinese authors (for details, see Appendix B). Chinese author names tend to be more difficult to disambiguate than other ethnic names (Kim & Diesner, 2016; Strotmann & Zhao, 2012). So, the ORCID-derived ground truth may provide optimistic performance results as Chinese names are disproportionally excluded from evaluation.

Despite such issues, this study is expected to motivate talented scholars to have interest in automatic labeling of training data for author name disambiguation. As more publications and new names enter digital libraries at an unprecedented rate (Bornmann & Mutz, 2015), automatic labeling can provide many practical solutions to supervised author name disambiguation for digital libraries. Identifying conditions of high-performing automatic labeling can benefit both scholars and stakeholders like academic institutions in need of unambiguous scholarly data for knowledge discovery and scholarly evaluation.

---

[16] http://dblp.org/faq/17334571

Acknowledgements

This work was supported by grants from the National Science Foundation (#1561687 and #1535370), the Alfred P. Sloan Foundation and the Ewing Marion Kauffman Foundation. We would like to thank anonymous reviewers for their helpful comments.

Appendix A: Construction of Self-Citation Relation

If a paper cites another paper, they are in citing-cited relation. From this paper-level citation information, scholars have constructed author-level citation relation. In Figure 6, Author A and Author B coauthors Paper 1, while Author C and Author D writes together Paper 2. If Paper 2 cites Paper 1 (paper-level citation), authors in Paper 2 are assumed to refer to authors in Paper 1. Thus, Author C is depicted to cite Author A and Author B, and Author D to cite Author A and Author B. If Author C is the same as Author A, they are in self-citation relation.

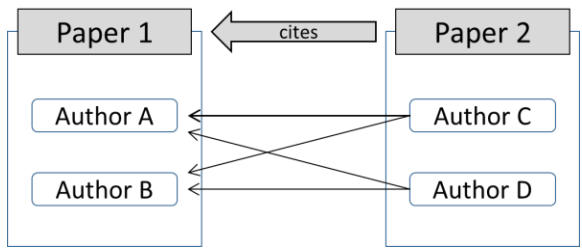

*Figure 6: An Illustration of Construction of Self-Citation Relation*

Appendix B: Representativeness Checks for ORCID-Linked Data

This section checks how the ORCID-linked data (Methodology > Data and Pre-processing > ORCID-Linkage) represent the whole data in this study. Following the method described in Representativeness Checks of Results, the ratios of name ethnicity and block size of ORCID-linked data are compared to those of the whole data. Figure 7 shows that in ORCID-linked data, Chinese names are under-represented while Hispanic and Italian names are over-represented while other ethnic names show similar ratios. This observation is contrasted to that from Figure 1 where Chinese names are slightly over-represented and English names are a little under-represented.

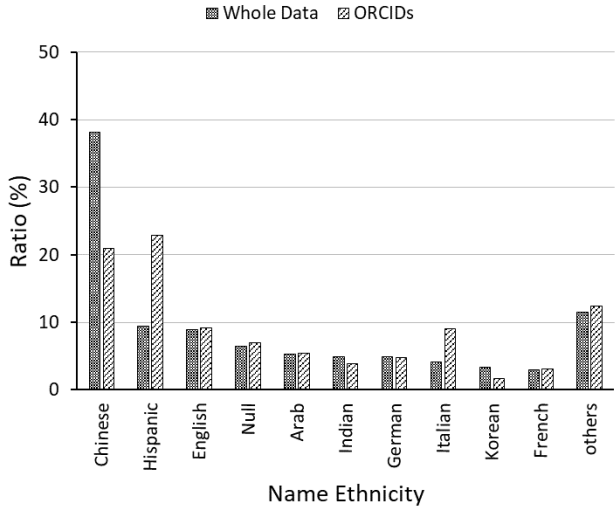

*Figure 7: Ratios of Name Ethnicity in ORCID-Linked Data (ORCIDs) Compared to Whole Data*

Regarding the block size distribution in Figure 8, the distribution plot of ORCIDs starts higher in y-axis (= ratio) than that of Random Data but falls below as x-value (= block size) increases. This means that ORCID-linked data contain more small blocks and less large blocks compared to randomly selected subset with the same number of name instances as ORCID-linked data, while automatically labeled data produce block size distribution quite similar to that of random data in Figure 2.

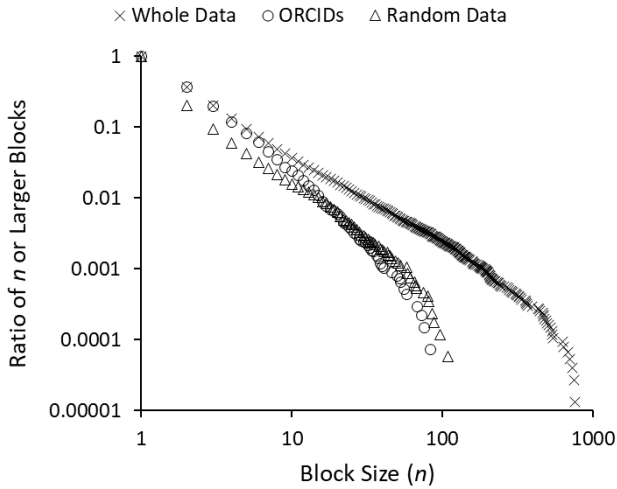

*Figure 8: Cumulative Ratios of Block Size on Log-Log Scale*